\documentclass[12pt]{iopart} 
\usepackage{iopams}
\usepackage[dvips]{graphicx}
\begin{document}
\title
{Interaction between  static holes in  a  quantum  dimer model on  the
kagome lattice}

\author{Gr\'egoire Misguich, Didina Serban and Vincent Pasquier}
\address{Service de Physique Th{\'e}orique,
CEA Saclay, 91191 Gif-sur-Yvette Cedex, FRANCE.\\
URA 2306 of CNRS}
\begin{abstract}

A quantum  dimer model (QDM) on the  kagome lattice with  an extensive
ground-state entropy was recently introduced [{\it Phys.  Rev. B} {\bf
67}, 214413 (2003)].  The ground-state energy of  this QDM in presence
of  one and  two  static  holes   is investigated  by  means  of exact
diagonalizations on lattices containing  up to 144  kagome sites.  The
interaction energy  between the  holes (at  distances up  to 7 lattice
spacings) is  evaluated  and  the   results  show no   indication   of
confinement at large hole separations.

\end{abstract}
\submitto{\JPCM} 


\newcommand{\La}{\line (1,0  ){12}}
\newcommand{\Lb}{\line (3,5 ){6}}
\newcommand{\Lc}{\line (-3,5 ){6}}
\newcommand{\Ld}{\line (-1,0){12}}
\newcommand{\Le}{\line (-3,-5){6}}
\newcommand{\Lf} {\line(3,-5){6}}
\newcommand{\C} {\circle*{4}}

\newcommand{\pA}{\put(-6,-10)}
\newcommand{\pB}{\put(6,-10)}
\newcommand{\pC}{\put(12,0)}
\newcommand{\pD}{\put(6,10)}
\newcommand{\pE}{\put(-6,10)}
\newcommand{\pF}{\put(-12,0)}
\newcommand{\pZ}{\put(0,0)}

\newcommand{\Hex}{\pA{\C}\pB{\C}\pC{\C}\pD{\C}\pE{\C}\pF{\C}}

\newcommand{\pG}{\put( 18,-10)}
\newcommand{\pH}{\put( 18,10)}
\newcommand{\pI}{\put(0,20)}
\newcommand{\pJ}{\put(-18,10)}
\newcommand{\pK}{\put(-18,-10)}
\newcommand{\pL}{\put(0,-20)}

\newcommand{\pM}{\put(24,20)}

\newcommand{\KagHex}{\pA{\C}\pB{\C}\pC{\C}\pD{\C}\pE{\C}\pF{\C}}
\newcommand{\KagStar}{\KagHex\pG{\C}\pH{\C}\pI{\C}\pJ{\C}\pK{\C}\pL{\C}}

\newcommand{\Z}{\mathbb{Z}_2}

\newcommand{\h}{\Omega}


\section{Introduction}

Quantum dimer models (QDM)  can provide some effective descriptions of
the  low-energy     singlet    dynamics    of    frustrated    quantum
antiferromagnets~\cite{rk88}.   A basis of  the Hilbert space of these
models is  made   by all (nearest-neighbor)  dimer  coverings   of the
lattice  and the  Hamiltonian allows  these  dimer to move along local
resonance  loops.  Two kinds  of phases are  well understood for these
models in  two dimensions:  dimer crystals~\cite{rk88} and  resonating
valence-bond    (RVB)  dimer liquids~\cite{ms01,msp02}.   Crystals are
characterized by  long-ranged dimer-dimer correlations and spontaneous
lattice symmetry breaking and RVB liquids  show no broken symmetry but
topological              order           and     $\mathbb{Z}_2$-vortex
excitations~\cite{rc89}. Importantly  also,  these two  phases can  be
distinguished by the behavior of holes (or spinons) when the system is
``doped'',  that is  we allow  sites   which are not  occupied  by any
dimer. Holes experience a mutual interaction which grows linearly with
their separation in a dimer crystal  and this interaction confine them
in pairs.   On the other hand, they  propagate independently  in a RVB
liquid background.  A first and  simple  step is  to consider QDM with
{\em static} holes (non-magnetic impurities or spinons).  In that case
a relevant  quantity is the ground-state energy  as a  function of the
hole  positions.  This  energy  goes to  a  constant when  two holes a
separated far  apart in a deconfined  system whereas it grows linearly
in   a    confined  system.\footnote{QDMs     at      Rokhsar-Kivelson
points~\cite{rk88} are an  exception: the  ground-state energy remains
exactly zero whatever the positions of the holes.}

\begin{table}
  \begin{center}
    \begin{tabular}{|c|c|cc|c|}
      \hline
	$\alpha$ & $n_\alpha$ & $|d_\alpha\rangle$  & $|\bar{d}_\alpha\rangle$ & $(-1)^{n_\alpha} $ \\
      \hline
      1 & 3 &
	\begin{picture}(50,34)(-24,-15)
	\pA{\La}\pC{\Lc}\pE{\Le}\KagHex
      \end{picture}&
      \begin{picture}(50,34)(-24,-15)
	\pB{\Lb}\pD{\Ld}\pF{\Lf}\KagHex
      \end{picture} & -1 \\
      \hline 
      $2,\cdots,4$ & 4 &\begin{picture}(50,34)(-24,-13)
	\pB{\La}\pC{\Lc}\pE{\Ld}\pF{\Lf}
	\KagHex\pG{\C}\pJ{\C}
      \end{picture}&
      \begin{picture}(50,34)(-24,-13)
	\pA{\La}\pG{\Lc}\pD{\Ld}\pJ{\Lf}
	\KagHex\pG{\C}\pJ{\C}
      \end{picture} & +1 \\
      $5,\cdots,10$ & 4 &\begin{picture}(50,34)(-24,-13)
	\pB{\La}\pC{\Lc}\pE{\Le}\pK{\La}
	\KagHex\pG{\C}\pK{\C}
      \end{picture}&
      \begin{picture}(50,34)(-24,-13)
	\pA{\La}\pG{\Lc}\pD{\Ld}\pF{\Le}
	\KagHex\pG{\C}\pK{\C}
      \end{picture} & +1 \\
      $11,\cdots,16$ & 4 &\begin{picture}(50,34)(-24,-13)
	\pB{\La}\pC{\Lb}\pD{\Ld}\pF{\Lf}
	\KagHex\pG{\C}\pH{\C}
      \end{picture}&
      \begin{picture}(50,34)(-24,-13)
	\pA{\La}\pG{\Lc}\pH{\Ld}\pE{\Le}
	\KagHex\pG{\C}\pH{\C}
      \end{picture}  & +1 \\
      \hline 
      $17,\cdots,19$ & 5 &\begin{picture}(50,34)(-24,-13)
	\pB{\La}\pC{\Lb}\pD{\Ld}\pJ{\Lf}\pK{\La}
	\KagHex\pG{\C}\pJ{\C}\pK{\C}\pH{\C}
      \end{picture}&
      \begin{picture}(50,34)(-24,-13)
	\pA{\La}\pG{\Lc}\pH{\Ld}\pE{\Ld}\pF{\Le}
	\KagHex\pG{\C}\pJ{\C}\pK{\C}\pH{\C}
      \end{picture}  & -1\\
      $20,\cdots,25$ & 5 &\begin{picture}(50,44)(-24,-23)
	\pC{\Lb}\pD{\Ld}\pJ{\Lf}\pK{\La}\pL{\Lb}
	\KagHex\pL{\C}\pJ{\C}\pK{\C}\pH{\C}
      \end{picture}&
      \begin{picture}(50,44)(-24,-23)
	\pA{\Lf}\pB{\Lb}\pH{\Ld}\pE{\Ld}\pF{\Le}
        \KagHex\pL{\C}\pJ{\C}\pK{\C}\pH{\C}
      \end{picture}  & -1\\
      $26,\cdots,31$ & 5 &\begin{picture}(50,44)(-24,-23)
	\pL{\Lb}\pG{\Lc}\pD{\Ld}\pJ{\Lf}\pK{\La}
	\KagHex\pL{\C}\pJ{\C}\pK{\C}\pG{\C}
      \end{picture}&
      \begin{picture}(50,44)(-24,-23)
	\pA{\Lf}\pB{\La}\pC{\Lc}\pE{\Ld}\pF{\Le}
	\KagHex\pL{\C}\pJ{\C}\pK{\C}\pG{\C}
      \end{picture}  & -1\\
      \hline 
      32 & 6 &\begin{picture}(50,50)(-24,-23)
	\pL{\Lb}\pG{\Lc}\pJ{\Lf}\pK{\La}\pH{\Ld}\pI{\Le}
	\KagStar
      \end{picture}&
      \begin{picture}(50,50)(-24,-23)
	\pA{\Lf}\pB{\La}\pC{\Lb}\pD{\Lc}\pE{\Ld}\pF{\Le}
	\KagStar
      \end{picture} & +1\\
      \hline 
    \end{tabular}
  \end{center}
  \caption[99]{The 8 different classes (up to rotations) of  dimerizations of an
    hexagon  of the kagome  lattice.}
  \label{tab:kloops}
\end{table}
We recently proposed a QDM on the kagome lattice which does not simply
fall in any of these two  categories~\cite{msp03}.  This model,  hereafter
called the $\mu$--model, was introduced from  the observation that the
dimer  kinetic energy terms arising  from  an overlap expansion of the
spin-$\frac{1}{2}$  Heisenberg  model~\cite{rk88} generally  have  non
trivial {\em signs} as soon as the competition of resonance loops with
{\em different lengths} is considered~\cite{ze95}. The Hamiltonian is:
\begin{equation}
	\mathcal{H}=-\sum_h \mu_h
\end{equation}
where 
\begin{equation}
	\mu_h= \sum_{\alpha=1}^{32}(-1)^{n_\alpha} 
	\left|d_\alpha(h) \right>\left< \bar{d}_\alpha(h)\right|
  	+
	{\rm H.c}
	\label{eq:muQDM}
\end{equation}
and $h$   runs   over      the    hexagons  of  the    lattice     and
$\left|d_\alpha(h)\right>$ is one  of the $32$ possible  dimerizations
of $h$  (table~\ref{tab:kloops}).    The sign  $n_\alpha$  counts  the
parity of the number of dimer involved\footnote{In the absence of that
sign the model  reduces to that  of reference~\cite{msp02}  and can be
solved exactly,  it has a  RVB  liquid ground-state  with  topological
order and gapped  $\mathbb Z_2$-vortex excitations.}.  It was realized
that such signs can lead to a new state, different from dimer crystals
or  RVB liquids.  In   our  previous study~\cite{msp03} the  following
results   were  obtained:   i)  The $\mu$--model    has   an extensive
ground-state  entropy  $\frac{1}{6}\log(2)$ per  kagome  site, that is
$50\%$ of  the  classical  dimer  entropy.  This  exponentially  large
degeneracy comes from a hidden, local, but non-abelian symmetry of the
model.  ii) It  is  possible to  choose  a basis of   the ground-state
manifold so  that dimer-dimer correlations   are short-ranged in  each
state.  These ground-states are thus dimer liquids.  iii) On the basis
of  exact  diagonalizations  we   argued  that, in   addition  to  the
ground-state degeneracy, the spectrum is likely to be gapless and that
energy-energy correlations (as well as susceptibilities) are likely to
be critical.

In the  present work we investigate  numerically the  effect of static
holes in the $\mu$--model.  This  issue is of particular importance as
Dommange {\it et al.}~\cite{dmnm03} pointed  out in a recent work that
static  holes  in    the  spin-$\frac{1}{2}$  kagome   antiferromagnet
experience a short-distance repulsion  and are probably  deconfined at
larger  distances.   Sindzingre  {\it  et  al.}\cite{slf01} previously
reached a similar conclusion  about {\em spinon deconfinement} from an
analysis of  the value of  the spin gap in a  24-site  sample with two
holes.   We show  in this  paper that  a somewhat  similar behavior is
observed in the $\mu$--QDM.

\section{$\mu$--model with holes}
\begin{figure}
	\begin{center}
	\includegraphics[width=5cm]{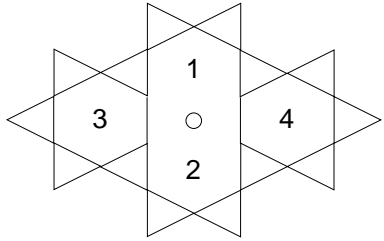}
	\caption{The hole forbids all resonance loops
	on hexagons 1 and 2 and suppresses some of the loops
	around 3 and 4.}\label{fig:hole}
	\end{center}
\end{figure}
As  any QDM,  the  $\mu$--model  can be   extended to  include  static
holes. These holes can equally represent  charge degrees of freedom or
neutral spinons  (unpaired  spin in  a   dimer background).   The new
Hamiltonian  $\mathcal{H}'$  contains    all the  kinetic  terms    of
$\mathcal{H}$  except those where  the resonance loop passes through a
missing site.  Consider  a hole which  belongs to two hexagons $1$ and
$2$ (figure~\ref{fig:hole}).  No loop of $\mu_1$ or $\mu_2$ survive in
$\mathcal{H}'$ because   they would   all pass   through  the  missing
site. As for hexagons 3 and 4, one half of  their resonance loops pass
through the hole and  must  be removed.  In  presence  of a hole   the
operators $\mu_3$  and $\mu_4$  thus  only contain 16  resonance loops
(instead   of 32).  However these  two  modified operators satisfy the
{\em    same  algebraic   relations~\cite{msp03}   as  the   hole-free
$\mu$'s}.  For  any hexagon  $h\ne1,2$   and for $i=3$   or  4 we have
$\mu_i^2=1$ and:
\begin{eqnarray}
     \mu_i\mu_h&=&\mu_h\mu_i
		\;\;\;\;\;\;\;\;\; h {\rm \;not\; neighbor\; of\;} i\\
      \mu_i\mu_h&=&-\mu_h\mu_i
			\;\;\;\;\; h {\rm \;neighbor\; of\;} i
			\label{eq:ac}
\end{eqnarray}
These relations  are   easy  to check   with the   help of  the  arrow
representation    of      dimer    coverings      of    the     kagome
lattice~\cite{ez93,msp02,msp03}.   It is also  easy  to check that the
argument  leading to an exponential degeneracy  $\sim  2^{N/6}$ of the
energy levels~\cite{msp03} holds even in the  presence of these static
holes. As a first result we thus  find that the extensive ground-state
entropy of the  $\mu$--model survives in the  presence of holes.  This
also allows to  use the  reduced  representation of  the Hilbert space
which was  used in reference~\cite{msp03}  to compute the  spectrum in
the absence  of  holes.   The   spectrum  is  non-degenerate in   this
representation, which has a   dimension  $\sim 2^{N/6}$   (instead  of
$2^{N/3+1}$ for the dimer Hilbert space).  The ground-state of systems
up to 48 hexagons  (144 kagome sites) can be  obtained with a standard
Lanczos algorithm. The  result was checked  (with and without holes)
against  direct  calculations  in  the dimer  Hilbert  space for small
systems ($N\leq48$).  We investigated samples with $N=36$, 48, 60, 72,
84,  108 and 144  kagome sites  ($N_h=$12, 16, 20,  24, 28,  36 and 48
hexagons).  Periodic  boundary conditions are  used and the  shapes of
these clusters are the same as those of reference~\cite{msp03}.

Interestingly this  representation allows to  compute the spectra of a
even system pierced by a {\em  single} hole. Strictly speaking the QDM
is  not defined  on such an  {\em  odd} sample but the  non-degenerate
representation  of   the $\mu$  algebra mentioned  above  can still be
constructed.    This trick  is  useful to   estimate  the  energy cost
$\Delta$ of a {\em single hole} in a given sample.

\section{Single hole ground-state energy}

\begin{figure}
	\begin{center}
	\includegraphics[width=8cm,trim=0 280 0 0 0,clip=true]{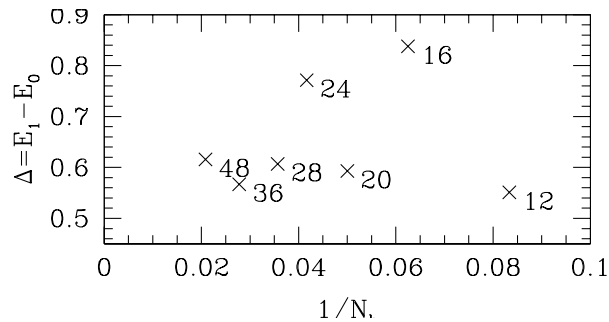}
	\caption{Energy cost $\Delta$ of a single hole in the $\mu$ QDM
	as a function of the number of hexagons $N_h$.}
	\label{fig:gap}
	\end{center}
\end{figure}
The ground-state energy in the absence of  holes is noted $E_0$, $E_1$
is the  energy with a single hole  (two neighboring  $\mu$'s removed).
If   $\langle\mu_i\mu_j\rangle$    correlations   are  neglected,  the
ground-state  energy would  increase   by $2\langle \mu  \rangle\simeq
0.88$  around each  hole (the ground-state  energy is  estimated to be
$\simeq     -0.44$    per      hexagon    in     the     thermodynamic
limit~\cite{msp03}). In fact  removing two neighboring $\mu$ operators
increases    the    ground-state   energy  by  $\Delta=E_1-E_0\sim0.6$
(figure~\ref{fig:gap}). It is easy  to understand why the actual  hole
gap $\Delta$ is smaller than the naive estimate above.  Because of the
anti-commutation  relations    between     nearby  $\mu$     operators
(equation~\ref{eq:ac}) the   system  cannot simultaneously  achieve  a
minimal energy ({\it  i.e}  $\mu=1$)   on two  neighboring   hexagons.
Removing some $\mu$  operators therefore decreases  the frustration on
their neighbors, which can acquire  in turn a larger expectation value
(lower energy).  This  larger ``polarization'' of  the hexagons around
the  hole  will enhance the  frustration on  their neighbors,  and the
corresponding $\mu$  will have  to reduce (slightly)  their expectation
value  compared with the bulk  value.  This mechanism produces spatial
oscillations in   $\langle\mu\rangle$  (data not  shown), oscillations
which have the same wave-vector as  the correlations which dominate in
the bulk~\cite{msp03}.

\section{Two-holes ground-state energy}

\begin{figure}
	\begin{center}
	\includegraphics[height=7cm]{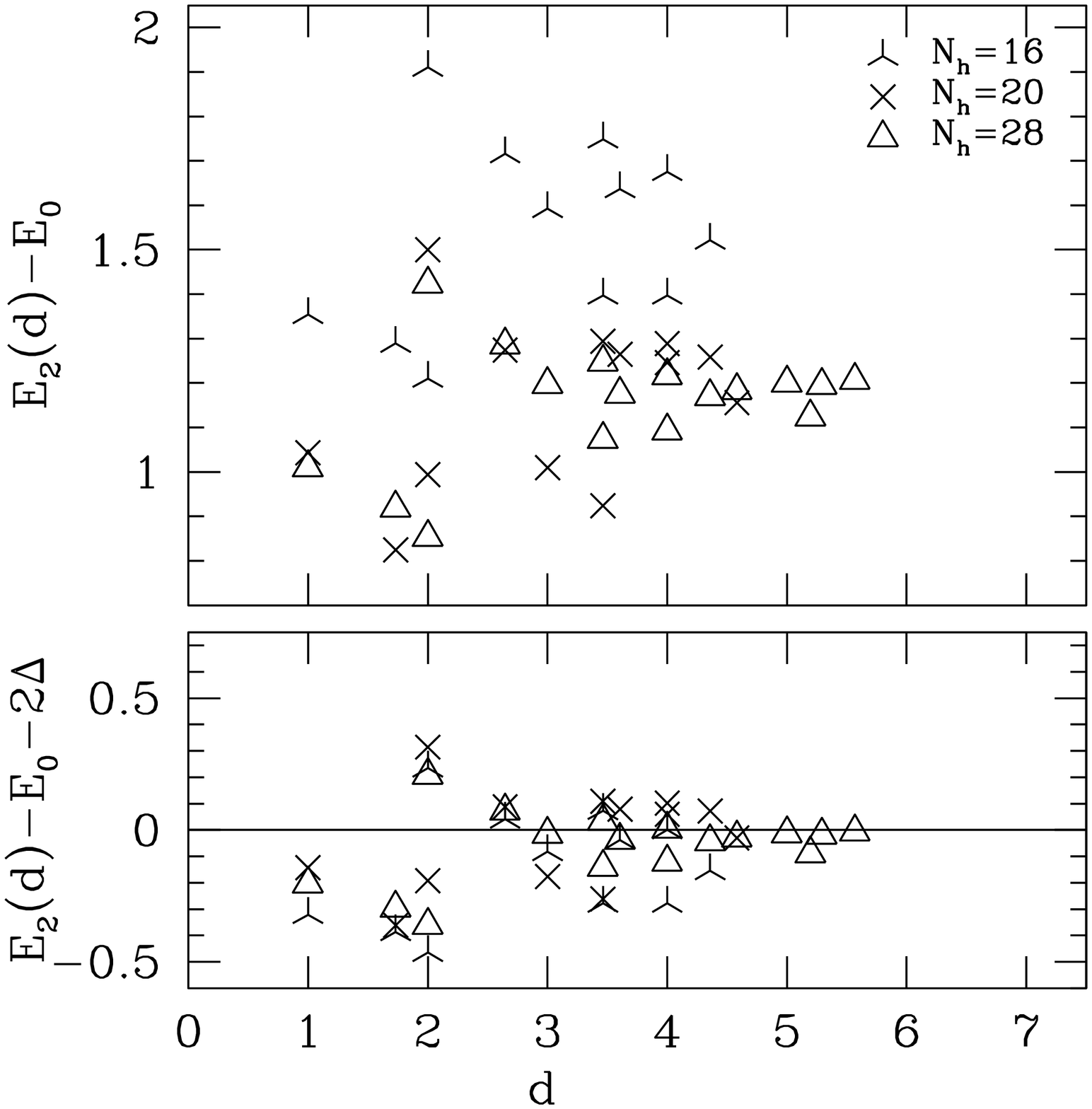}
	\includegraphics[height=7cm,trim=80 0 0 0]{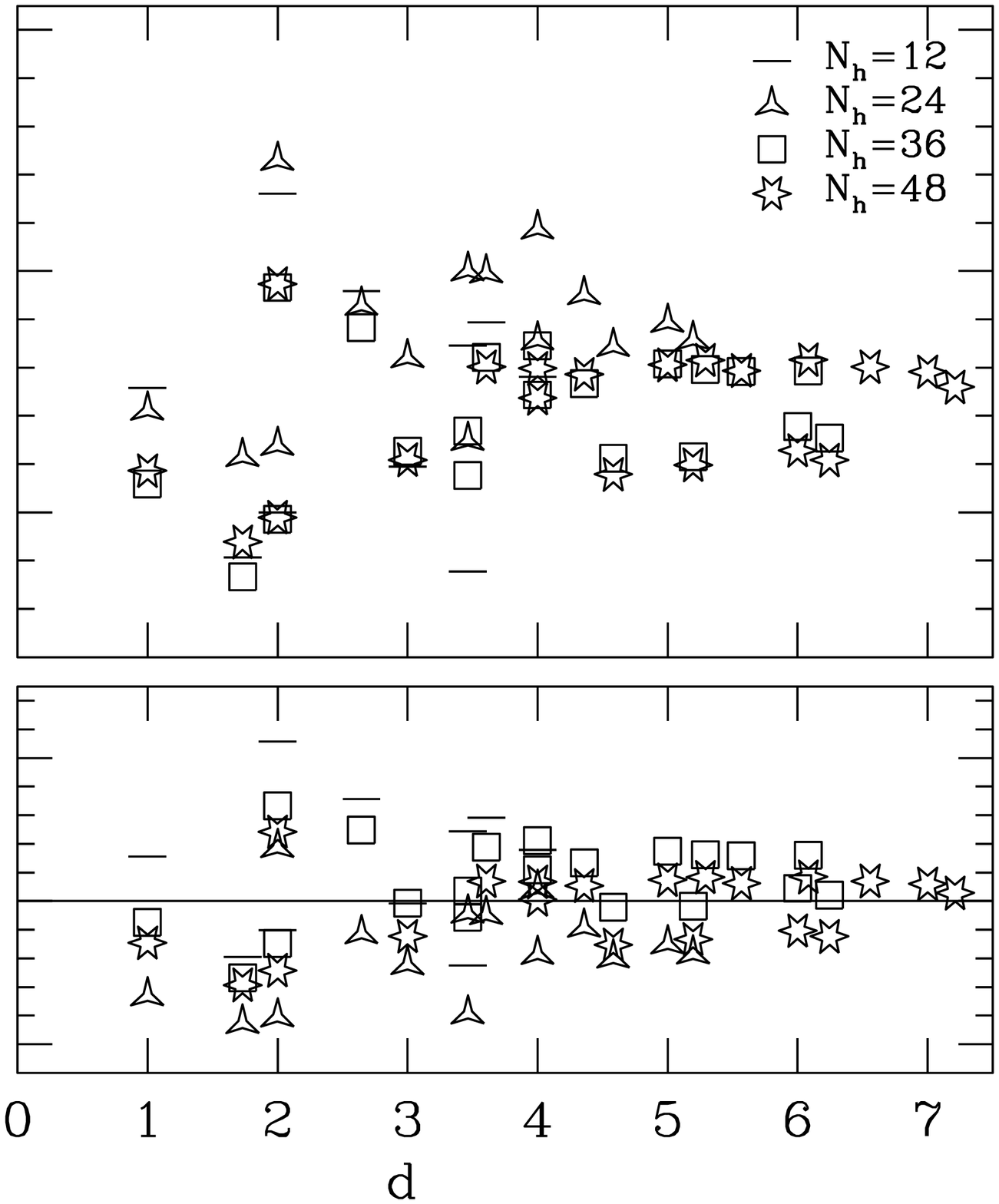}
	\caption{Energy $E_2(d)$ of the ground-state of the $\mu$--QDM
	in presence of two holes at distance $d$.
	In the top panels this energy is compared to the energy $E_0$
	of the system without holes and in the lower panels
	$E_2(d)$ is compared to the hole-free energy
	$E_0$ corrected by twice the energy cost $\Delta=E_1-E_0$ of a
	single hole.}
	\label{fig:2holes}
	\end{center}
\end{figure}

The difference between the energy $E_2(d)$  with two holes at distance
$d$     and    the  energy $E_0$     without      holes  is shown   in
figure~\ref{fig:2holes}.   In  the analysis   of  Ref.~\cite{msp03} it
appeared    that    the     $\mu$-model    has     significant   local
$\langle\mu_i\mu_j\rangle$   correlations  with   a  period  of  three
hexagons.  It is therefore convenient to  plot separately the data for
$N_h$ not multiple of three and the others ($N_p=12$,  24, 36 and 48),
which do not frustrate  the local order.   At short distance, when the
two holes belong   to a common  hexagon,  only 3 $\mu$  operators  are
removed from $\mathcal{H}$ and the energy cost is roughly $E_2-E_0\sim
\frac{3}{2}\Delta$.  This happens for   $d=1$, $\sqrt{3}$ as well  as
for $d=2$ when the two holes are on opposite sites of an hexagon. This
is   a short-distance effect     because for $d>2$  the  number  $\mu$
suppressed is always four.  In other words, when  two holes sit on the
same hexagon they minimize the number of  loops which are ``lost'' for
resonances~\footnote{This  effect  can also  be   found, to a  smaller
extent, in the two-hole  energies of the spin-1/2 model~\cite{dmnm03}:
Comparing the two ways two holes can be  at distance $d=2$, the energy
is  always lower  when they belong  to  the  same hexagon.  There  is,
however, no strong reduction of $E_2$ for  $d=1$ and $\sqrt{3}$ in the
spin model as we have in  the $\mu$--QDM.}.  At intermediate distances
($2\leq  d\leq\sqrt{12}$) the energy   decreases  with distance in   a
regular  way for all samples.   In this range of   $d$ the behavior is
thus  reminiscent of the strong hole  repulsion observed in the kagome
Heisenberg model by Dommange {\it et al.}~\cite{dmnm03}.

For $d\geq4$  the data suggest  that the  energy  $E_2(d)$ goes  to  a
constant. The values are indeed close to the energy $E_0+2\Delta$ (see
lower panels of   figure~\ref{fig:2holes}) which  is expected if   the
dimer  background was   not   mediating any  interaction  between  the
holes. We therefore argue that the $\mu$ model is not confining static
holes.  It is interesting to note   that in the   samples which do not
frustrate the local  order  (right panels of  figure~\ref{fig:2holes})
the   hole-hole interaction seems  to  decay more slowly with distance
than in the other samples.  This rather slow decay is not incompatible
with the interesting suggestion~\cite{dmnm03} of a $1/d$ behavior.  In
addition, weak oscillations that can be observed.  They may be related
to the  oscillations of $\langle\mu\rangle$  mentioned in the previous
section and  to  the  (presumably) quasi  long-ranged correlations  in
$\langle\mu_i\mu_j\rangle^c$ discussed in Ref.~\cite{msp03}



\section*{References}
\begin{thebibliography}{99}

\bibitem{rk88}
Rokhsar D S and Kivelson S A 1988 {\it Phys. Rev. Lett.} {\bf 61},  2376


\bibitem{ms01}
Moessner R and Sondhi S L 2001 {\it Phys. Rev. Lett.} {\bf 86} 1881

\bibitem{msp02}
Misguich G, Serban  D and Pasquier V 2002  {\it Phys. Rev. Lett.} {\bf
89} 137202

\bibitem{rc89}
Read and Chakraborty 1989 {\it Phys. Rev.} B {\bf 40} 7133



\bibitem{msp03}
Misguich G, Serban  D and Pasquier V 2003  {\it Phys. Rev.} B {\bf 67} 214413

\bibitem{ze95}
Elser V and Zeng C 1995 {\it Phys. Rev.} B {\bf 51} 8318

\bibitem{slf01}
Sindzingre  P, Lhuillier C  and   Fouet J.-B.  2002, {\em Advances  in
Quantum Many-Body Theory} {\bf 6} 90, cond-mat/0110283

\bibitem{dmnm03}
Dommange S, Mambrini M, Normand B and  Mila F {\it Preprint} cond-mat/0306299

\bibitem{ez93}
Elser V and Zeng C 1993 {\it Phys. Rev.} B {\bf 48} 13647

\end{thebibliography}
\end{document}